\documentclass{Interspeech}

\interspeechcameraready 

\title{Fine-Tuning ASR for Stuttered Speech: \\ Personalized vs. Generalized Approaches}

\author[]{Dena}{Mujtaba}
\author[]{Nihar R.}{Mahapatra}
\affiliation[nocounter]{}{Michigan State University}{USA}
\email{\{mujtabad,nrm\}@msu.edu}


\usepackage{comment}
\usepackage{multirow}
\usepackage{booktabs} 

\begin{document}

\maketitle

\begin{abstract}
Stuttering---characterized by involuntary disfluencies such as blocks, prolongations, and repetitions---is often misinterpreted by automatic speech recognition (ASR) systems, resulting in elevated word error rates and making voice-driven technologies inaccessible to people who stutter. The variability of disfluencies across speakers and contexts further complicates ASR training, compounded by limited annotated stuttered speech data. In this paper, we investigate fine-tuning ASRs for stuttered speech, comparing generalized models (trained across multiple speakers) to personalized models tailored to individual speech characteristics. Using a diverse range of voice-AI scenarios, including virtual assistants and video interviews, we evaluate how personalization affects transcription accuracy. Our findings show that personalized ASRs significantly reduce word error rates, especially in spontaneous speech, highlighting the potential of tailored models for more inclusive voice technologies.
    
\end{abstract}

\section{Introduction}
Voice AI, systems driven by speech input, have become ubiquitous in contemporary life due to recent significant performance improvements. However, these technologies remain largely inaccessible to people who stutter, who constitute approximately 1\% of the global population \cite{yairi2013epidemiology}. Such individuals encounter substantial barriers in routine activities involving virtual assistants, automated video captioning, and voice-controlled devices \cite{mujtaba2024lost,lea2023user,li2024re}. This limited accessibility stems chiefly from inaccuracies in automatic speech recognition (ASR) models when processing speech containing \textit{disfluencies}—unintended disruptions like repetitions, sound prolongations, and interjections. Indeed, prior studies have demonstrated substantial disparities in ASR word error rates between stuttered and fluent speech \cite{mujtaba2024lost}, highlighting systemic biases. These inaccuracies do not merely inconvenience individuals who stutter; they exacerbate social marginalization and impede equitable access to automated services and employment opportunities \cite{mujtaba2019ethical,gerlach2018stuttering,plexico2019influence}.

\subsection{Related Work}

A significant obstacle to improving ASR accuracy for stuttered speech is the paucity of annotated speech data \cite{mujtaba2024lost,macdonald2021disordered}. Effective datasets must be sufficiently extensive and reflective of the diverse manifestations of stuttering across different speakers and speaking contexts \cite{tichenor2021variability}. Historically, much of the research on ASR for stuttering has concentrated on disfluency identification—classifying and filtering disfluencies, or employing predicted disfluency probabilities as supplementary inputs for ASR \cite{changawala24_interspeech,lea2021sep}. While datasets such as SEP-28k \cite{lea2021sep}, and methods including Whister \cite{changawala24_interspeech} and the work by Shonibare et al. \cite{shonibare2022enhancing} have been proposed, they inadequately address inherent inaccuracies in ASR systems. Errors from disfluency detection can propagate into subsequent transcription tasks. Additionally, attempts to distinctly categorize disfluencies often blur the lines between stuttered and non-stuttered speech \cite{einarsdottir2005have}. Complicating matters further, overlapping and concurrent disfluencies, characteristic of conditions like cluttering \cite{st2003cluttering}, have yet to be effectively resolved by existing systems. Thus, enhancing ASR accuracy directly through model training remains crucial.

Recently, some studies have addressed data scarcity issues by fine-tuning existing ASR models on stuttered speech. Lea et al. \cite{lea2023user} fine-tuned parameters within Apple's ASR framework using speech data from individuals who stutter. Similarly, Müller et al. \cite{muller2024hypernetworks} employed parameter-efficient fine-tuning strategies aimed at atypical speech, specifically focusing on stuttering. Furthermore, Google's Project Euphonia aggregated stuttered speech data but primarily targeted other speech conditions, such as dysarthria, that differ significantly from stuttering in disfluency types and variability \cite{macdonald2021disordered,tobin2022assessing,tichenor2021variability}.

\subsection{Our Contributions}

This paper directly addresses these gaps by developing and evaluating several fine-tuned ASR models designed specifically to enhance performance on stuttered speech. We systematically explore personalized ASR—tailored to individual speech characteristics—and generalized ASR, trained across multiple speakers, hypothesizing that personalized approaches yield greater accuracy given the variability in stuttering patterns. Furthermore, we evaluate transcription accuracy in distinct contexts (spontaneous versus read speech) and across diverse voice AI application scenarios, spanning domains such as healthcare, finance, and virtual assistants.

In summary, our key contributions are: \textbf{(1) Fine-Tuning a Large Pre-trained ASR for Stuttered Speech:} We investigate the effectiveness of fine-tuning Whisper, a robust pre-trained encoder-decoder ASR model, using limited stuttered speech datasets. To train these models and conduct various ablation studies, we employ parameter-efficient fine-tuning and low-rank adaptation (LoRA) \cite{hu2021lora}, which stabilizes training and mitigates overfitting risks inherent in small datasets due to fewer parameter updates. We evaluate our models using word error rate (WER) and character error rate (CER), with the goal of improving performance on disfluent speech by enhancing accuracy, and investigating granular transcription errors specifically due to repeated syllables or sound prolongations. \textbf{(2) Personalized versus Generalized ASR Comparison:} We present an in-depth comparison between personalized ASR models—tailored specifically to an individual’s speech—and generalized ASR models, which currently represent the standard approach. Although generalized models are designed to be broadly applicable across multiple speakers, we systematically explore the feasibility and advantages of personalization by examining the amount of training data required and assessing model performance across diverse speech contexts. Our objective is to determine the most effective strategies for developing ASR systems that provide equitable and accessible voice technology for people who stutter. \textbf{(3) ASR Performance Across Different Contexts:} We assess ASR performance and the impact of fine-tuning across multiple voice AI contexts, explicitly including both spontaneous and read speech. For this purpose, we utilize two datasets. First, we employ FluencyBank \cite{ratner2018fluency}, a dataset containing interview and reading samples from people who stutter, widely used in stuttering research but limited in size and scope. To enhance the diversity of our study, we aggregate a new dataset of stuttered speech collected from 18 individuals through job interview–related questions (spontaneous speech) and read speech prompts from various voice AI scenarios. This combined approach enables us to evaluate how ASR models trained on one type of speech generalize to the other, thus effectively capturing the variability inherent in stuttering. Through this work, we aim to advance the development of accessible ASR systems for people who stutter and others with speech variations.

\section{Methodology}

\subsection{Automatic Speech Recognition Model: Whisper}
We utilize Whisper~\cite{radford2023robust}, a sequence-to-sequence transformer-based model capable of multilingual and multitask speech transcription. Whisper processes audio in 30-second segments via its encoder, producing a sequential representation used by its decoder for transcription. Trained on over 680k hours of diverse speech data, Whisper effectively manages noisy audio and various speech conditions. Additionally, its training objective includes secondary tasks such as language identification, significantly enhancing transcription quality compared to earlier ASR systems. Despite these advantages, Whisper has a notable limitation—it occasionally hallucinates, generating nonexistent words or phrases in its transcriptions, particularly during long conversational pauses~\cite{mujtaba2024lost}. Such hallucinations and related errors frequently occur with stuttered speech. Prior research indicates Whisper Large's WER on stuttered speech is approximately twice that on fluent speech~\cite{mujtaba2024lost}. Although Whisper achieves a comparatively low overall WER for fluent speech, this disparity in performance persists when evaluating various disfluency types commonly associated with stuttering~\cite{mujtaba2024lost}.

\subsection{Datasets}
\subsubsection{FluencyBank}
We evaluated ASR performance using two datasets. The first, FluencyBank (FB) \cite{ratner2018fluency}, contains videos of individuals who stutter, captured during interviews and while reading passages. Each video includes transcripts in the Codes for Human Analysis of Transcripts (CHAT) format, a standard speech-language pathology annotation protocol accurately representing disfluencies and segmenting audio into individual utterances (short phrases or sentences)~\cite{mujtaba2024lost}. Annotation involved multiple transcribers independently transcribing each file using the CLAN software, a tool specifically designed for preparing CHAT transcripts~\cite{macwhinney2017tools}, followed by steps to ensure inter-annotator agreement. FluencyBank comprises speech from 12 participants, totaling 1,373 utterances and 2.21 hours of audio.

\begin{figure}[t]
\centering
\includegraphics[width=.99\columnwidth,trim={0cm 0cm 0cm 0cm},clip]{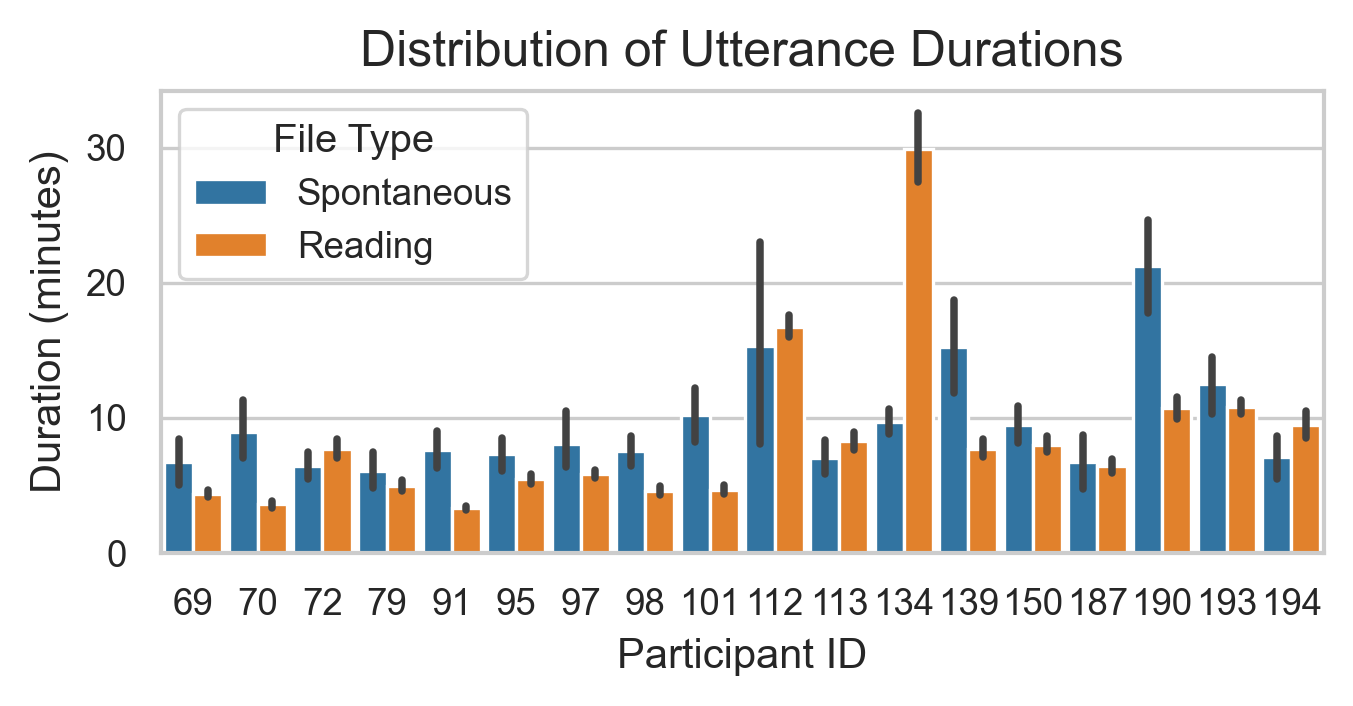}
\caption{Duration of recorded audio per participant in the HeardAI (HAI) dataset. Participants are assigned random IDs to preserve anonymity.}
\label{fig:participants}
\vspace*{-.3in}
\end{figure}

\subsubsection{HeardAI Dataset}
The second dataset, which we term HeardAI (HAI), is a newly aggregated collection of stuttered speech designed to address the limitations of FluencyBank—namely, its limited contextual diversity and overall size. HeardAI encompasses 18 different categories of voice AI usage, designed by a team of researchers specializing in AI, speech-language pathology, and psychology, as shown in Figure \ref{fig:prompt_category}.

To collect speech samples, we conducted a study in which participants were prompted with various phrases from voice AI categories. These prompts were randomly selected from a predefined database—developed by our research team—that contains 50 unique prompts per voice-AI category and captures commonly used phrases such as ``Lock the front door'' and ``Turn on closed captioning.'' Each participant recorded 60 read prompts, followed by five spontaneous prompts in which they answered job interview–related questions. We chose a job interview context to help reduce accuracy disparities in employment settings—an area where individuals who stutter have historically faced discrimination and lower job satisfaction \cite{gerlach2018stuttering,plexico2019influence}.

Once all recordings were completed, they were transcribed by a team of speech-language pathologists (SLPs) with expertise in stuttering and speech transcription. The transcription team annotated the speech samples using CLAN, following steps to ensure inter-annotator agreement. Read prompts were verified for accuracy, and any problematic recordings were excluded from analysis. In total, data from 18 participants were included, comprising 2.51 hours of read speech (1,071 utterances) and 1.05 hours of spontaneous speech (405 utterances). Figure \ref{fig:participants} shows the distribution of recorded minutes per participant, averaging 8.4 minutes each. Due to participant privacy concerns, the dataset cannot be publicly released at this time. We use this dataset to conduct a fine-grained analysis of ASR performance and fine-tuning effectiveness across different speech types (read versus spontaneous), voice AI categories, and among individuals who stutter.

\subsection{Parameter-Efficient Fine-Tuning}

Given the substantial parameter counts of large pre-trained models (e.g., Whisper large variant has 1.5B parameters~\cite{radford2023robust}), comprehensive fine-tuning of such models often exceeds single-GPU resource limits. To efficiently conduct personalization experiments across numerous participants—and to utilize larger model architectures—we employ parameter-efficient fine-tuning techniques that adapt large pre-trained models to new tasks or datasets by updating only a small subset of parameters~\cite{hu2021lora}. One such technique is low-rank adaptation (LoRA), which replaces updates to selected weight parameters with trainable low-rank matrix representations while keeping the remaining pre-trained weights frozen. This approach reduces the number of trainable parameters as well as the computational resources and time required for training~\cite{hu2021lora}.

Specifically, given a pre‐trained weight matrix $W \in \mathbb{R}^{d \times k}$,
its adapted representation is defined as:
\[
W_0 + \Delta W = W_0 + BA,
\]
where \(W_0\) denotes the frozen pre‐trained weights, and \(\Delta W = BA\) represents the low‐rank update. Here, we use matrices $B \in \mathbb{R}^{d \times r} \quad$ and $\quad A \in \mathbb{R}^{r \times k}$, with rank \(r \ll \min(d,k)\)~\cite{hu2021lora}. Additionally, a scaling factor \(\alpha\) is introduced to modulate the magnitude of the update:
$
\Delta W = \frac{\alpha}{r}\, BA.
$
During training, the original weights \(W_0\) remain frozen, receiving no gradient updates; only the matrices \(A\) and \(B\) are updated. Consequently, the forward pass is computed as: $
h = W_0 x + BAx$, where \(x\) denotes the input. This method significantly reduces the number of trainable parameters and storage requirements, as only the low-rank updates \(\Delta W\) need to be saved.

In this work, we employ LoRA with rank \(r = 128\), scaling factor \(\alpha = 256\), and dropout \(p = 0.05\) on all linear layers of Whisper, settings previously determined to be effective~\cite{muller2024hypernetworks}. This approach enables training multiple personalized models efficiently, even with limited computational resources.

\subsection{Evaluation Metrics}

To effectively assess performance, we employ two metrics—each offering a unique perspective on transcription accuracy for stuttered speech. The first metric is \textit{word error rate} (WER), a standard measure in ASR evaluation, defined as $WER = \frac{S + D + I}{S + D + C}$,
where \(S\), \(D\), and \(I\) denote the number of word substitutions, deletions, and insertions, respectively, and \(C\) denotes the number of correct words. WER provides an overall measure of transcript inaccuracy at the word level. The second metric is \textit{character error rate} (CER), which evaluates inaccuracy at the character level, capturing finer-grained transcription errors. Such granularity is particularly important for analyzing stuttered speech, where repetitions, partial words, or individual sound-level errors frequently occur.

\subsection{Experiments and Implementation}
For our study, we fine-tuned \textit{Whisper small} (weights from HuggingFace\footnote{\url{https://huggingface.co/openai/whisper-small}}) through five experiments. In \textbf{Experiment 1}, we trained a generalized model using FluencyBank and evaluated its performance on our HAI dataset, establishing a baseline with a combination of spontaneous and read speech. Then, in \textbf{Experiment 2}, we investigated how much additional participant-specific training data is required to improve personalized ASR performance. Specifically, we divided each participant's 60 read speech prompts from the HAI dataset into five training folds of 12 prompts each and one evaluation fold (12 prompts). We incrementally trained five models per participant (using 0, 12, 24, 36, and 48 training prompts) and evaluated them on the held-out fold to study how WER scales with training dataset size.

Subsequent experiments focused on personalized ASR fine-tuning, starting from the model obtained in Experiment 1. In \textbf{Experiment 3}, we fine-tuned individual ASR models for each participant using a three-fold cross-validation strategy applied to their read prompts, split by prompt category to prevent voice AI context overlap. These models were evaluated on the held-out read prompts as well as spontaneous speech samples to examine the generalization from read to spontaneous speech. Similarly, in \textbf{Experiment 4}, we fine-tuned personalized ASR models using only the spontaneous speech samples and evaluated them on read prompts to analyze generalization from spontaneous to read contexts. Finally, in \textbf{Experiment 5}, we trained personalized models for each participant using three-fold cross-validation on a combination of read and spontaneous speech, with no prompt category overlap, to examine the overall effect of combining both speech types on transcription performance.

The fine-tuning process for all experiments was conducted on a single NVIDIA V100 GPU. For all experiments except Experiment 2, we trained models for 20 epochs using a learning rate of $\gamma = 0.00001$, a warmup period of 400 steps, and a batch size of 8 with gradient accumulation applied every other step. For Experiment 2, we employed similar hyperparameters but reduced the number of epochs to 7 and the warmup period to 50 steps, as this experiment focused on evaluating how quickly ASR performance improved with increasing dataset size rather than achieving optimal final accuracy. Additionally, to standardize the input data, we applied normalization using the Whisper normalizer from HuggingFace~\cite{normalizer}, which adjusts casing and punctuation prior to evaluation.

\begin{figure}[t]
\centering
\caption{Word error rate (WER) per prompt category for Whisper-small (Original), generalized ASR from Experiment 1 (E1), and personalized ASR from Experiment 5 (E5).}
\vspace*{-.1in}
\includegraphics[width=\columnwidth,trim={0cm 0cm 0cm 0cm},clip]{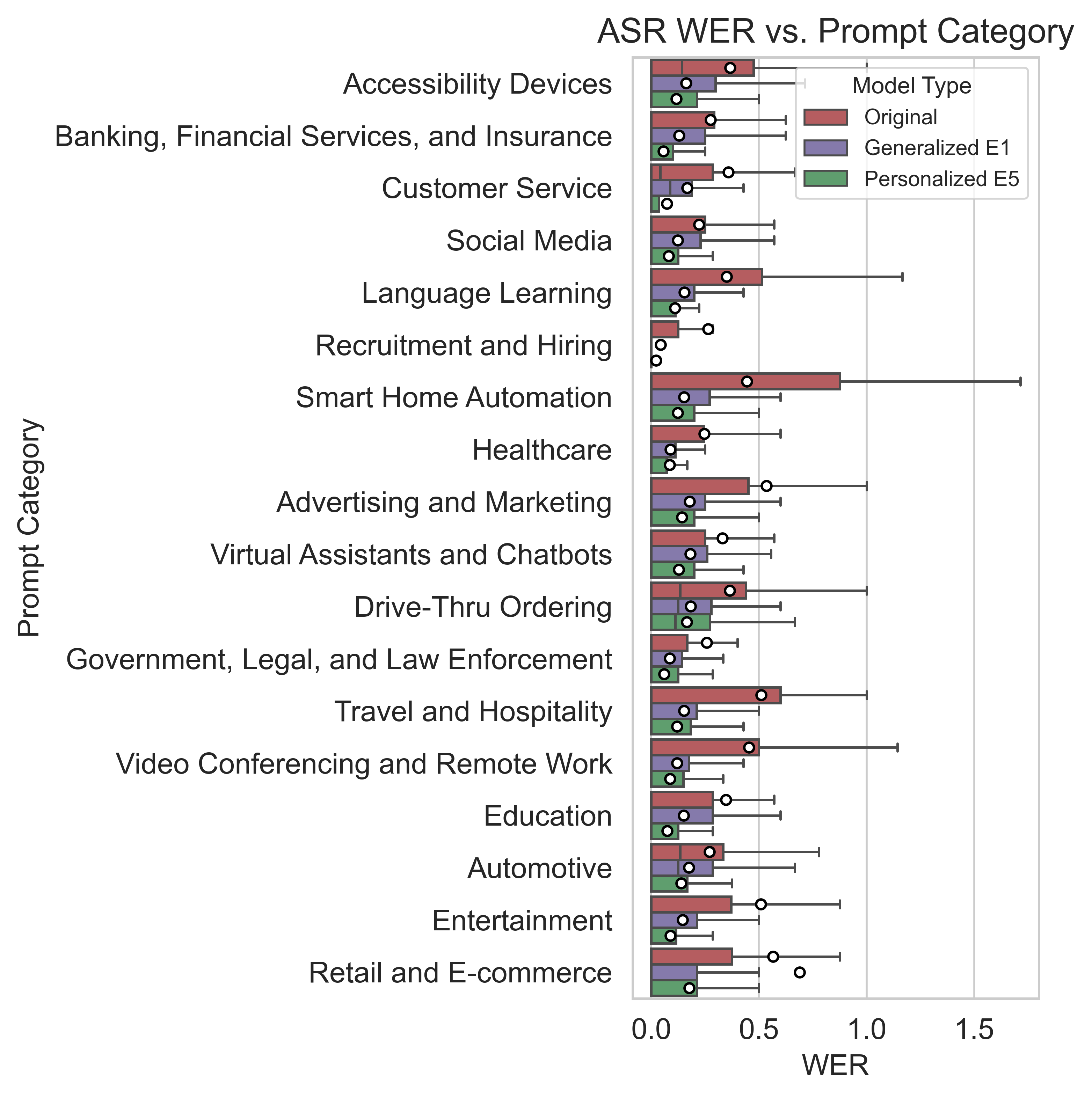}
\label{fig:prompt_category}
\vspace*{-.4in}
\end{figure}

\section{Results and Discussion}

\begin{table}[t]
    \centering
        \caption{Word error rate (WER) and character error rate (CER) results for Whisper-small (WS) and fine-tuned ASR variations evaluated on the HeardAI dataset (reading and spontaneous speech samples). ``Gen-E1'' indicates the generalized model from Experiment~1, while ``WS-P'' denotes personalized models from Experiments~3--5 (Section~2.5). For personalized ASRs, we report mean results across participants.}
    \begin{tabular}{l|l|l|l|l}
        \hline
        \multirow{2}{*}{\textbf{ASR}} & \multicolumn{2}{c|}{\textbf{Reading}}  & \multicolumn{2}{c}{\textbf{Spontaneous}} \\ \cline{2-5} 
         & \textbf{WER} & \textbf{CER}& \textbf{WER} & \textbf{CER} \\ \specialrule{1pt}{0pt}{0pt}
WS	& 0.3388	& 0.2151	& 0.2217	& 0.1540 \\ 
Gen-E1 & 0.1612	& 0.0909	& 0.1839	& 0.0877 \\ 
WS-P-E3& 0.1355	& 0.0826	& 0.3186	& 0.2164 \\ 
WS-P-E4& 0.1550	& 0.0881	& -	& -\\ 
WS-P-E5& 0.0939	& 0.0478	& 0.0981	& 0.0573\\ 
        \hline 
    \end{tabular}
    \label{tbl:overview}
\end{table}

\begin{figure}[t]
\centering
\includegraphics[width=.9\columnwidth,trim={0cm 0cm 0cm 0cm},clip]{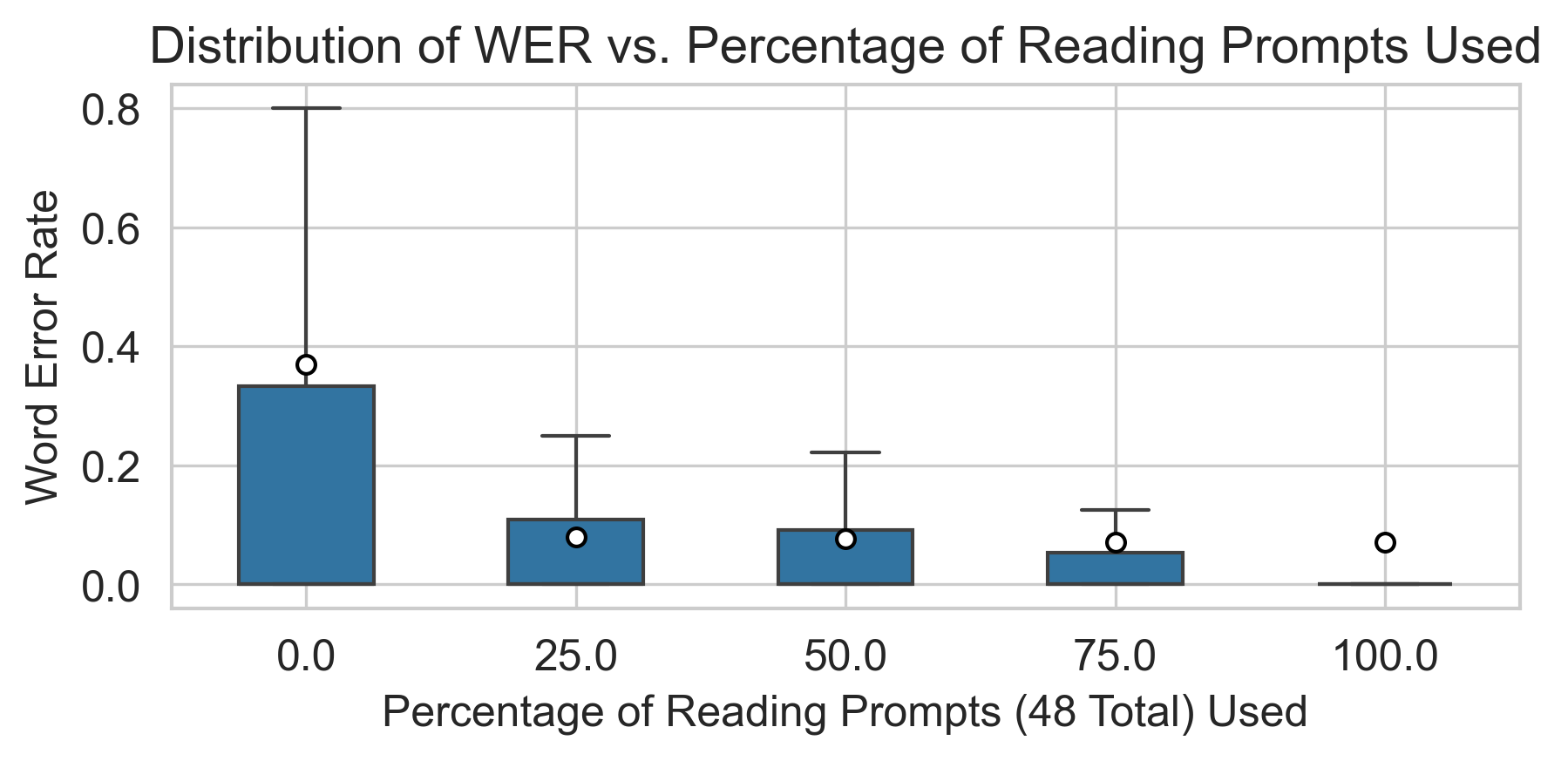}
\caption{Word error rate (WER) as a function of the percentage of reading prompts used for participant-specific training, averaged across HeardAI participants. White circles indicate mean values; outliers are omitted for clarity.}
\label{fig:study}
\vspace*{-.2in}
\end{figure}

In Table~\ref{tbl:overview}, we present the results for the untrained Whisper-small model on the HAI dataset. Both WER and CER are notably high for read and spontaneous speech samples. Figure~\ref{fig:prompt_category} further illustrates these results by prompt category, highlighting that certain categories, notably ``Smart Home Automation,'' ``Language Learning,'' ``Travel and Hopitality,'' and Video Conferencing and Remote Work'' exhibit comparatively higher WER. These elevated error rates may be due in part to factors such as longer or more complex prompts within these categories. Table~\ref{tbl:overview} also compares performance across fine-tuned ASR models from our experiments. Fine-tuning on FluencyBank significantly improved transcription accuracy, reducing WER on read speech from 33.9\% to 16.12\% and spontaneous speech from 22.17\% to 18.39\%. Figure~\ref{fig:prompt_category} demonstrates that this improvement is consistent across all prompt categories.

Next, in Experiment~2, we analyze how much participant-specific speech data is required to improve ASR accuracy through fine-tuning. As illustrated in Figure~\ref{fig:study}, there is a substantial reduction in WER after only 25\% of the reading prompts are included in training. While accuracy continues to improve as more data is added, mean WER improvements become marginal beyond 50\% of the prompts. This indicates that a relatively small amount of audio data may be sufficient for effectively fine-tuning ASR models to transcribe stuttered speech accurately.

\begin{figure}[t]
\centering
\includegraphics[width=.99\columnwidth,trim={0cm 0cm 0cm 0cm},clip]{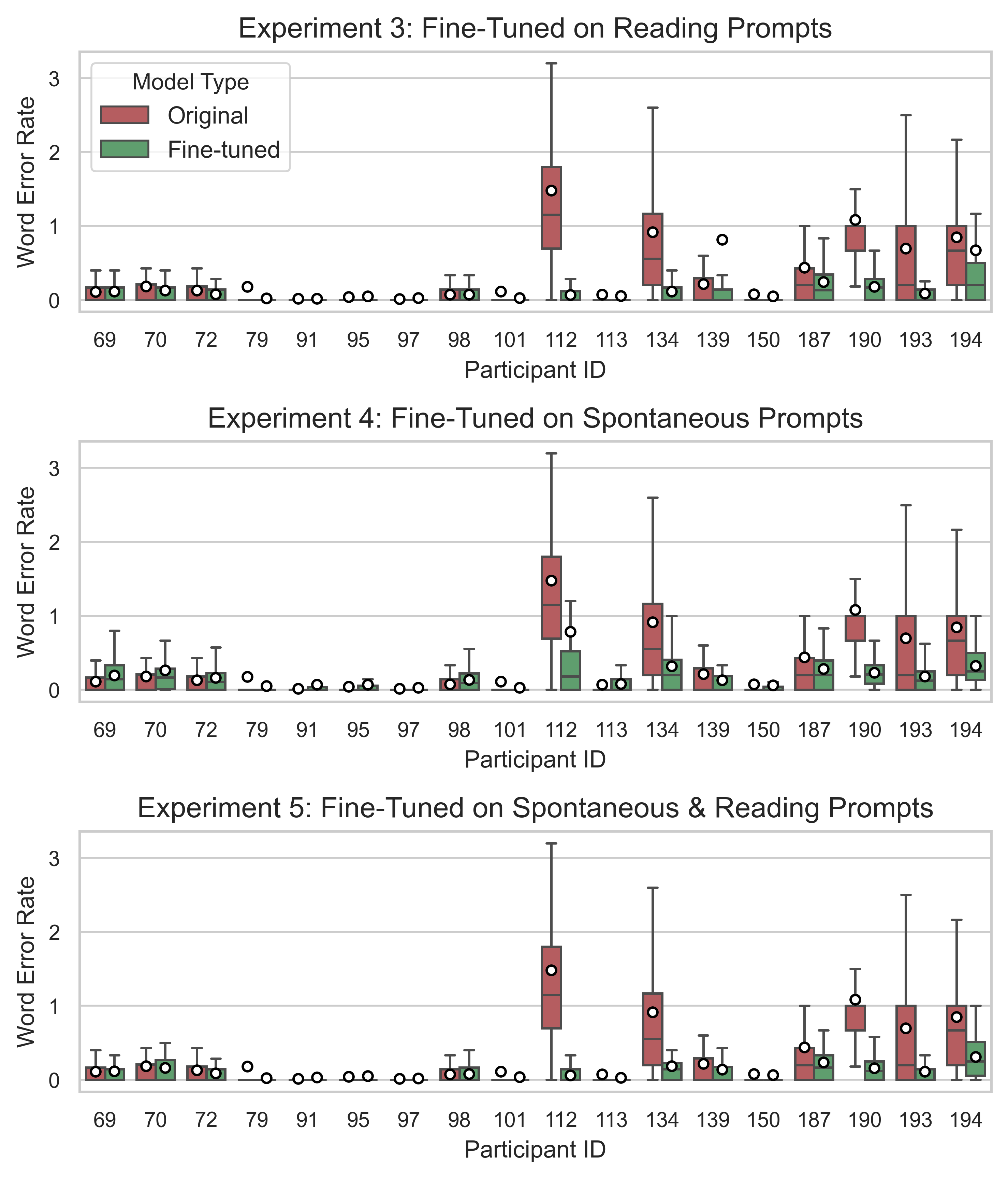}
\caption{Word error rate (WER) per participant on reading prompts from the HeardAI dataset, comparing personalized fine-tuning experiments (Experiments~3--5) with the original Whisper-small ASR model (no fine-tuning). Participants are assigned random IDs to maintain anonymity.}
\label{fig:results_dist}
\vspace*{-.2in}
\end{figure}

Last, results from Experiments~3--5 are summarized in Table~\ref{tbl:overview} and Figure~\ref{fig:results_dist}. We find that even fine-tuning on small amounts of participant-specific speech significantly reduces both WER and CER, consistently outperforming the generalized ASR model. Notably, the personalized ASR from Experiment~5 achieves the best performance, reducing WER on read speech from 16.12\% (generalized model) to 9.39\%, and from 18.39\% to 9.81\% on spontaneous speech—approaching accuracy levels seen with non-stuttered speech~\cite{mujtaba2024lost}. Interestingly, the personalized ASR trained solely on spontaneous samples (Experiment~4), despite using fewer training samples compared to read speech, achieved performance comparable to Experiment~3, underscoring the substantial contribution of spontaneous speech to ASR improvement. This may reflect the observation that stuttering typically occurs more frequently in spontaneous speech than in read speech~\cite{rasskazov2007so}. Overall, the findings from Experiment~5 clearly demonstrate the advantage of personalized ASR models over generalized approaches. As illustrated in Figure~\ref{fig:results_dist}, personalized fine-tuning markedly reduces WER variability across participants, with improvements also consistent across prompt categories (Figure~\ref{fig:prompt_category}). Notably, participants with higher initial WER—potentially indicative of more pronounced stuttering—experienced proportionally larger accuracy gains.

\section{Conclusion}
In this paper, we investigated the performance differences between generalized and personalized ASR approaches for stuttered speech, specifically through fine-tuning large pre-trained models. Using a newly aggregated dataset covering diverse voice-AI scenarios, we analyzed how transcription accuracy varied across different applications and speech contexts among individuals who stutter. Our findings demonstrate that fine-tuning ASR models using even small amounts of participant-specific speech data significantly improves transcription accuracy compared to generalized models. Additionally, we found that spontaneous speech contributes more substantially to model improvement than read speech, reflecting the inherently higher variability and complexity of spontaneous speech in stuttering. Looking ahead, we plan to extend this research by examining specific disfluency patterns in greater detail and further modeling the distinct speech characteristics of people who stutter. Ultimately, our goal is to advance ASR technologies toward greater accessibility, fairness, and inclusivity, benefiting not only people who stutter but also the broader population, as nearly everyone exhibits some degree of speech disfluency.

\ifinterspeechfinal
\section{Acknowledgements}
This material is based upon work supported by the U.S. National Science Foundation under Grant No.~2345086. We thank the HeardAI team members for their assistance with dataset collection and transcription.
\fi

\bibliographystyle{IEEEtran}
\bibliography{refs}

\begin{thebibliography}{10}
\providecommand{\url}[1]{#1}
\csname url@samestyle\endcsname
\providecommand{\newblock}{\relax}
\providecommand{\bibinfo}[2]{#2}
\providecommand{\BIBentrySTDinterwordspacing}{\spaceskip=0pt\relax}
\providecommand{\BIBentryALTinterwordstretchfactor}{4}
\providecommand{\BIBentryALTinterwordspacing}{\spaceskip=\fontdimen2\font plus
\BIBentryALTinterwordstretchfactor\fontdimen3\font minus
  \fontdimen4\font\relax}
\providecommand{\BIBforeignlanguage}[2]{{%
\expandafter\ifx\csname l@#1\endcsname\relax
\typeout{** WARNING: IEEEtran.bst: No hyphenation pattern has been}%
\typeout{** loaded for the language `#1'. Using the pattern for}%
\typeout{** the default language instead.}%
\else
\language=\csname l@#1\endcsname
\fi
#2}}
\providecommand{\BIBdecl}{\relax}
\BIBdecl

\bibitem{yairi2013epidemiology}
E.~Yairi and N.~Ambrose, ``Epidemiology of stuttering: 21st century advances,''
  \emph{Journal of fluency disorders}, vol.~38, no.~2, pp. 66--87, 2013.

\bibitem{mujtaba2024lost}
D.~Mujtaba, N.~Mahapatra, M.~Arney, J.~Yaruss, H.~Gerlach-Houck, C.~Herring,
  and J.~Bin, ``Lost in transcription: Identifying and quantifying the accuracy
  biases of automatic speech recognition systems against disfluent speech,'' in
  \emph{Proceedings of the 2024 Conference of the North American Chapter of the
  Association for Computational Linguistics: Human Language Technologies
  (Volume 1: Long Papers)}, 2024, pp. 4795--4809.

\bibitem{lea2023user}
C.~Lea, Z.~Huang, J.~Narain, L.~Tooley, D.~Yee, D.~T. Tran, P.~Georgiou, J.~P.
  Bigham, and L.~Findlater, ``From user perceptions to technical improvement:
  Enabling people who stutter to better use speech recognition,'' in
  \emph{Proceedings of the 2023 CHI Conference on Human Factors in Computing
  Systems}, 2023, pp. 1--16.

\bibitem{li2024re}
J.~Li, S.~Wu, and G.~Leshed, ``Re-envisioning remote meetings: Co-designing
  inclusive and empowering videoconferencing with people who stutter,'' in
  \emph{Proceedings of the 2024 ACM Designing Interactive Systems Conference},
  2024, pp. 1926--1941.

\bibitem{mujtaba2019ethical}
D.~F. Mujtaba and N.~R. Mahapatra, ``Ethical considerations in {AI}-based
  recruitment,'' in \emph{2019 IEEE International Symposium on Technology and
  Society (ISTAS)}.\hskip 1em plus 0.5em minus 0.4em\relax IEEE, 2019, pp.
  1--7.

\bibitem{gerlach2018stuttering}
H.~Gerlach, E.~Totty, A.~Subramanian, and P.~Zebrowski, ``Stuttering and labor
  market outcomes in the {United States},'' \emph{Journal of Speech, Language,
  and Hearing Research}, vol.~61, no.~7, pp. 1649--1663, 2018.

\bibitem{plexico2019influence}
L.~W. Plexico, M.-B. Hamilton, H.~Hawkins, and S.~Erath, ``The influence of
  workplace discrimination and vigilance on job satisfaction with people who
  stutter,'' \emph{Journal of Fluency Disorders}, vol.~62, p. 105725, 2019.

\bibitem{macdonald2021disordered}
R.~L. MacDonald, P.-P. Jiang, J.~Cattiau, R.~Heywood, R.~Cave, K.~Seaver, M.~A.
  Ladewig, J.~Tobin, M.~P. Brenner, P.~C. Nelson \emph{et~al.}, ``Disordered
  speech data collection: Lessons learned at 1 million utterances from {Project
  Euphonia}.'' in \emph{Interspeech}, vol. 2021, 2021, pp. 4833--4837.

\bibitem{tichenor2021variability}
S.~E. Tichenor and J.~S. Yaruss, ``Variability of stuttering: Behavior and
  impact,'' \emph{American Journal of Speech-Language Pathology}, vol.~30,
  no.~1, pp. 75--88, 2021.

\bibitem{changawala24_interspeech}
V.~Changawala and F.~Rudzicz, ``Whister: Using {Whisper’s} representations
  for stuttering detection,'' in \emph{Interspeech 2024}, 2024, pp. 897--901.

\bibitem{lea2021sep}
C.~Lea, V.~Mitra, A.~Joshi, S.~Kajarekar, and J.~P. Bigham, ``{SEP}-28k: A
  dataset for stuttering event detection from podcasts with people who
  stutter,'' in \emph{ICASSP 2021-2021 IEEE International Conference on
  Acoustics, Speech and Signal Processing (ICASSP)}.\hskip 1em plus 0.5em minus
  0.4em\relax IEEE, 2021, pp. 6798--6802.

\bibitem{shonibare2022enhancing}
O.~Shonibare, X.~Tong, and V.~Ravichandran, ``Enhancing {ASR} for stuttered
  speech with limited data using detect and pass,'' \emph{arXiv preprint
  arXiv:2202.05396}, 2022.

\bibitem{einarsdottir2005have}
J.~Einarsd{\'o}ttir and R.~J. Ingham, ``Have disfluency-type measures
  contributed to the understanding and treatment of developmental stuttering?''
  2005.

\bibitem{st2003cluttering}
K.~O. St.~Louis, L.~J. Raphael, F.~L. Myers, and K.~Bakker, ``Cluttering
  updated,'' \emph{The ASHA Leader}, vol.~8, no.~21, pp. 4--22, 2003.

\bibitem{muller2024hypernetworks}
M.~M{\"u}ller-Eberstein, D.~Yee, K.~Yang, G.~V. Mantena, and C.~Lea,
  ``Hypernetworks for personalizing {ASR} to atypical speech,''
  \emph{Transactions of the Association for Computational Linguistics},
  vol.~12, pp. 1182--1196, 2024.

\bibitem{tobin2022assessing}
J.~Tobin, Q.~Li, S.~Venugopalan, K.~Seaver, R.~Cave, and K.~Tomanek,
  ``Assessing {ASR} model quality on disordered speech using {BERTScore},''
  \emph{arXiv preprint arXiv:2209.10591}, 2022.

\bibitem{hu2021lora}
E.~J. Hu, Y.~Shen, P.~Wallis, Z.~Allen-Zhu, Y.~Li, S.~Wang, L.~Wang, and
  W.~Chen, ``{LoRA}: Low-rank adaptation of large language models,''
  \emph{arXiv preprint arXiv:2106.09685}, 2021.

\bibitem{ratner2018fluency}
N.~B. Ratner and B.~MacWhinney, ``Fluency{Bank}: A new resource for fluency
  research and practice,'' \emph{Journal of fluency disorders}, vol.~56, pp.
  69--80, 2018.

\bibitem{radford2023robust}
A.~Radford, J.~W. Kim, T.~Xu, G.~Brockman, C.~McLeavey, and I.~Sutskever,
  ``Robust speech recognition via large-scale weak supervision,'' in
  \emph{International Conference on Machine Learning}.\hskip 1em plus 0.5em
  minus 0.4em\relax PMLR, 2023, pp. 28\,492--28\,518.

\bibitem{macwhinney2017tools}
B.~MacWhinney, ``Tools for analyzing talk part 1: The {CHAT} transcription
  format,'' \emph{Carnegie.[Google Scholar]}, vol.~16, 2017.

\bibitem{normalizer}
HuggingFace, ``Normalization and pre-tokenization,''
  \url{https://huggingface.co/learn/nlp-course/chapter6/4}.

\bibitem{rasskazov2007so}
I.~Rasskazov and N.~Rasskazova, ``Why do so many stutterers fail to stutter
  when alone and how can this phenomonen be used in treatment,'' in
  \emph{International Stuttering Awareness Day Online Conference}, 2007.

\end{thebibliography}

\end{document}